\documentclass[12pt,showpacs,showkeys,amsmath,amssymb]{revtex4}
\usepackage{amsmath,amsfonts,amsthm,amscd,amssymb,latexsym}
\usepackage{bm}
\usepackage{dcolumn}
\usepackage{graphicx}
\usepackage{epstopdf}
\usepackage{color}
\usepackage{epsf}
\usepackage{epsfig}
\usepackage{graphicx, epic, eepic, color}

\newcommand{\beq}{\begin{equation}}
\newcommand{\eeq}{\end{equation}}
\providecommand{\abs}[1]{\lvert#1\rvert}

\begin{document}

\title{Twisted Gravitational Waves}

\author{Donato \surname{Bini}$^{1,2}$}
\email{donato.bini@gmail.com}
\author{Carmen \surname{Chicone}$^{3,4}$}
\email{chiconec@missouri.edu}
\author{Bahram \surname{Mashhoon}$^{4,5}$}
\email{mashhoonb@missouri.edu}

\affiliation{
$^1$Istituto per le Applicazioni del Calcolo ``M. Picone'', CNR, I-00185 Rome, Italy\\
$^2$ICRANet, Piazza della Repubblica 10, I-65122 Pescara, Italy\\
$^3$Department of Mathematics, University of Missouri, Columbia, Missouri 65211, USA\\
$^4$Department of Physics and Astronomy, University of Missouri, Columbia, Missouri 65211, USA\\
$^5$School of Astronomy, Institute for Research in Fundamental
Sciences (IPM), P. O. Box 19395-5531, Tehran, Iran\\
}

\date{\today}

\begin{abstract}
In general relativity (GR), linearized gravitational waves propagating in empty Minkowski spacetime along a fixed spatial direction have the property that the wave front is the Euclidean plane.  Beyond the linear regime, exact plane waves in GR have been studied theoretically for a long time and many exact vacuum solutions of the gravitational field equations  are known that represent \emph{plane}  gravitational waves. These have parallel rays and uniform wave fronts. It turns out, however, that GR also admits exact solutions representing gravitational waves propagating along a fixed direction that are \emph{nonplanar}. The wave front is then nonuniform and the bundle of rays is twisted. We find a class of solutions representing nonplanar  unidirectional gravitational waves and study some of the properties of these twisted waves. 
\end{abstract}


\maketitle

\section{Introduction}

It is a natural consequence of a field theory of universal gravitation that gravitational radiation should exist in some form. The recent detection of gravitational waves is consistent with Einstein's theory of gravitation; see, for instance, Ref.~\cite{GW} and the references cited therein. This circumstance provides further motivation to study exact gravitational waves within the context of general relativity (GR). 

The present paper is devoted to exact \emph{unidirectional} \emph{nonplanar} gravitational waves. 
Indeed, exact solutions of GR corresponding to gravitational \emph{pp}-waves and plane waves have received much attention~\cite{Brink, BPR, JBG, R1, Griffiths:2009dfa}. Many exact solutions also exist for the case of two colliding gravitational plane waves, as summarized in Ref.~\cite{JBG}, where the main result seems to be the formation of a spacetime singularity as a consequence of the collision itself, i.e. of the mutual interaction and focusing of the two waves.
Finally, special solutions corresponding to impulsive plane waves with  Dirac delta-function profiles have also been studied~\cite{Khan:1971vh}.

Beyond unidirectional waves, we note that the emission of gravitational radiation by an isolated astrophysical system generally results in a nearly spherical expanding wave front far away from the source. In this connection, exact solutions of GR include the Robinson-Trautman expanding spherical  gravitational waves, the nonexpanding gravitational waves of the Kundt class with spherical or hyperbolic wave fronts depending upon the sign of the cosmological constant $\Lambda$, and their impulsive analogs~\cite{RT, BP1, BP2}. Other exact solutions include the Einstein-Rosen cylindrical gravitational waves~\cite{AE, MQ}. Detailed treatments of exact solutions of GR are contained in Refs.~\cite{R1, Griffiths:2009dfa}. 

Plane gravitational waves possess five Killing vector fields and thus form a subclass of the plane-fronted waves with parallel rays known as the \emph{pp}-waves. A characteristic feature of the \emph{pp}-waves in general relativity is that they admit a covariantly constant null vector field $k$, 
\beq \label{I1}
k_{\mu ; \nu} = 0\,, \qquad k_\mu\,k^\mu = 0\,, 
\eeq  
where $k$ is the Killing vector field representing the propagation vector of the
 \emph{pp}-wave~\cite{R1,Griffiths:2009dfa}. Let us recall that \emph{any} gravitational wave propagating along a fixed direction
 in space---say, the $z$ direction---always admits a null Killing propagation vector $k= \partial_t + \partial_z$, 
\beq \label{I2}
k_\mu\,k^\mu = 0\,, \qquad  k_{\mu ; \nu} + k_{\nu ; \mu} = 0\,, \qquad k_{\mu ; \nu}\,k^{\nu} = 0\,,
\eeq
which therefore  represents a nonexpanding and shearfree null geodesic congruence. Furthermore, if the congruence is twistfree as well, then we recover Eq.~\eqref{I1} and we have a \emph{pp}-wave. It is however possible that the congruence is such that 
\beq \label{I3}
 \mathbb{T}_{\mu \nu} =k_{[\mu ; \nu]} = k_{[\mu , \nu]} \ne 0\,,
\eeq
in which case we have a twisted gravitational wave, namely, an exact unidirectional nonplanar gravitational wave in GR.  Here, $\mathbb{T}_{\mu \nu}$ is the antisymmetric \emph{twist tensor}. The wave propagation vector $k$ is orthogonal to the wave front, which are surfaces of constant $u$. Thus, 
\beq \label{I3a}
\omega^2 :=  \frac{1}{2}\, k_{[\mu; \nu]}\,k^{[\mu; \nu]} = \frac{1}{2}\, \mathbb{T}_{\mu \nu}\,\mathbb{T}^{\mu \nu} = 0\,, 
\eeq
which is consistent with the notion that the \emph{twist scalar} $\omega$ vanishes if and only if the null geodesic congruence is hypersurface-orthogonal. Therefore, $k_\mu = \partial u/\partial x^\mu$ for \emph{plane} gravitational waves, while $k_\mu = \Omega(x)\,\partial u/\partial x^\mu$ with $\partial \Omega/\partial x^\mu  \ne 0$ for \emph{nonplanar} gravitational waves under consideration in this paper.

An interesting solution which can be associated with a nonplanar gravitational wave propagating along a fixed direction in space is due to B. K. Harrison~\cite{Harrison:1959zz}. 
We encountered this solution within the context of \emph{cosmic jets} and studied the peculiar asymptotic behavior of its timelike geodesics in our previous work~\cite{Bini:2017qnd}. The purpose of the present paper is to present a class of  exact solutions of GR corresponding to twisted gravitational waves, i.e. waves propagating along a given direction (the $z$ axis, say) with wave fronts that are nonplanar and hence nonuniform. 

In section II, we find a class of twisted gravitational waves with wave fronts that have negative Gaussian curvature. This class contains Harrison's solution~\cite{Harrison:1959zz} as a special case, as discussed in section III. In section IV, we study the geodesics of a new member of this class and show that timelike and null geodesics behave as in the Harrison solution~\cite{Bini:2017qnd}. How can one observationally distinguish twisted gravitational waves from plane waves? Linearized gravitational waves in GR are \emph{transverse}, a property that is present in exact plane waves as well. On the other hand, a twisted gravitational wave exhibits a \emph{longitudinal} feature in addition to its transverse property. Can this longitudinal character be an observational signature of twisted gravitational waves?  We address this problem in section V by comparing the solutions of the Jacobi equation in two related situations involving plane and twisted gravitational waves. Section VI contains a discussion of our results.

\section{A Class of Exact Gravitational Waves}

In our recent work on \emph{cosmic jets}~\cite{Bini:2017qnd}, we considered, among other exact solutions of GR, 
 a Ricci-flat solution  due to Harrison~\cite{Harrison:1959zz} that we wrote as
\begin{equation}\label{G1}
ds^2 = -\check{x}^{4/3}\, dt^2 + \lambda^2\, \check{u}^{6/5}\,dx^2+ \check{x}^{-2/3}\,\check{u}^{-2/5}\,dy^2 + \check{x}^{4/3}\,dz^2\,,
\end{equation}  
where
\begin{equation}\label{G2}
\check{x} := \frac{x}{T_0}\,, \qquad \check{u} := \frac{t-z}{T_0}\,.
\end{equation} 
We use units such that $G=c=1$ throughout this paper; moreover, the signature of the spacetime metric is +2 and greek indices run from 0 to 3, while latin indices run from 1 to 3. Here, $T_0$ is a constant length and $\lambda > 0$ is a dimensionless parameter.  The dimensionless quantity $\check{u}$ is simply proportional to the retarded null coordinate $u = t-z$. Equation~\eqref{G1} is related by means of straightforward coordinate transformations to the second degenerate solution that Harrison classified as  the ``III-2; $D_2$" metric~\cite{Harrison:1959zz}. This spacetime is of type \emph{D} in the Petrov classification~\cite{DIRC}. The parameter $\lambda$ can be set equal to unity by rescaling the spacetime coordinates. Harrison's spacetime has curvature singularities at $w = 0$, where $w := \check{x}^{1/3}\,\check{u}^{1/5}$. 

Harrison's solution describes an exact gravitational wave (GW) propagating in the $z$ direction. The wave front is the $(x, y)$ surface with metric 
\begin{equation}\label{G3}
d\sigma^2= dx^2 +  \check{x}^{-2/3}\,dy^2\,,
\end{equation} 
which has negative Gaussian curvature. It thus represents an exact unidirectional nonplanar gravitational radiation field; that is, a \emph{twisted gravitational wave} (TGW).

To investigate such TGW solutions, we consider a metric of the form
\begin{equation}\label{G4}
 ds^2 = - e^{A(x, u)}(dt^2-dz^2) + e^{B(x, u)}\,dx^2 + e^{C(x, u)}\,dy^2\,,
\end{equation}
where we assume for the sake of simplicity that all lengths are measured in units of $T_0=1$. The field equations for exact gravitational waves, namely, $R_{\mu \nu} = \Lambda\, g_{\mu \nu}$, have been worked out for the case of metric~\eqref{G4} in Appendix A. The Ricci-flat field equations in this case can be obtained from Eqs.~\eqref{A3}--\eqref{A7} given in Appendix A by setting the cosmological constant $\Lambda$ equal to zero, namely,  
\beq \label{G5}
A_x^2+2\,A_x\,C_x =0\,,
\eeq
where $A_x:=\partial A/\partial x$, etc., 
\beq \label{G6}
2\,A_{xx} +A_x^2 -A_x(B_x+C_x)=0\,,
\eeq
\beq \label{G7}
2\,C_{xx} +C_x^2 -A_x^2 -B_x\,C_x=0\,,
\eeq
\beq \label{G8}
A_{xu} + C_{xu} = \frac{1}{2}\, A_x(B_u+C_u) + \frac{1}{2}\, C_x(B_u-C_u)\,
\eeq
and
\beq \label{G9}
A_u\, (B_u+C_u)  = \frac{1}{2}\,(B_u^2+C_u^2) + B_{uu}+C_{uu}\,.
\eeq

In the analysis of these equations, it proves useful to introduce 
\beq \label{G10}
\mathcal{A} = e^A\,, \qquad \mathcal{B} = e^B\,,\qquad \mathcal{C} = e^C\,.
\eeq
For instance, in terms of these new functions, Eq.~\eqref{G9} takes the form
\beq \label{G11}
\frac{\mathcal{A}_u}{\mathcal{A}}\, \left(\frac{\mathcal{B}_u}{\mathcal{B}}+\frac{\mathcal{C}_u}{\mathcal{C}}\right)  = -\frac{1}{2}\,\left(\frac{\mathcal{B}_u^2}{\mathcal{B}^2}+\frac{\mathcal{C}_u^2}{\mathcal{C}^2}\right) + \frac{\mathcal{B}_{uu}}{\mathcal{B}}+\frac{\mathcal{C}_{uu}}{\mathcal{C}}\,.
\eeq

It follows from Eq.~\eqref{G5} that either  $A_x=0$ in case A or  $A_x + 2\,C_x = 0$ in case B. These cases will be investigated in turn in the rest of this section.

\subsection{$A_x=0$}

This condition implies that $A$ is only a function of $u$. Since $dt^2-dz^2 = du dv$, where $v = t+z$ is the advanced null coordinate, we can absorb $\mathcal{A}$ in the redefinition of the retarded null coordinate. Thus one may, in effect, set $A = 0$. Then, Eqs.~\eqref{G7} and~\eqref{G8} can be written as 
\beq \label{G12}
B_x-C_x = 2\, \frac{C_{xx}}{C_x}\,, \qquad   B_u-C_u = 2\,\frac{C_{xu}}{C_x}\,,
\eeq
which can be simply integrated and the result is
\beq \label{G13}
B = C + 2\, \ln{C_x} + 2\, \ln \lambda\,,
\eeq
where $\lambda > 0$ is an integration constant. The problem thus reduces to finding solutions of the system
\beq \label{G14}
\mathcal{B} = \lambda^2\, \frac{\mathcal{C}_x^2}{\mathcal{C}}\,, \qquad  \frac{1}{2}\,\left(\frac{\mathcal{B}_u^2}{\mathcal{B}^2}+\frac{\mathcal{C}_u^2}{\mathcal{C}^2}\right) = \frac{\mathcal{B}_{uu}}{\mathcal{B}}+\frac{\mathcal{C}_{uu}}{\mathcal{C}}\,.
\eeq

It is simple to find a solution using separation of variables. That is, let
\beq \label{G15}
\mathcal{C} = X(x)\,U(u)\,.
\eeq
Then, it follows from Eq.~\eqref{G14} that $2\,UU_{uu} - U_u^2=0$, so that 
\beq \label{G16}
U(u)  =  (a u+b)^2\,,
\eeq
where $a$ and $b$ are integration constants and the metric in this case takes the form
\begin{equation}\label{G17}
 ds^2 = - dt^2 + (au+b)^2\left(\lambda^2\, \frac{1}{X}dX\,^2 + X\,dy^2\right) + dz^2\,,
\end{equation}
which represents \emph{flat} spacetime. We note that 
\begin{equation}\label{G18}
  \lambda^2\, \frac{1}{X}dX\,^2 + X\,dy^2 = dr^2 + r^2\,d\theta^2 \,,
\end{equation}
where $X = r^2/(4\lambda^2)$ and $y= 2\lambda\, \theta$. This case is further discussed in Appendix B. 

Let us now return to system~\eqref{G14} and introduce the change of variables
\beq \label{G18a}
\mathcal{P} = \sqrt{\mathcal{B}}\,, \qquad   \mathcal{Q} = \sqrt{\mathcal{C}}\,,
\eeq
since $\mathcal{B}$ and $\mathcal{C}$ are positive by definition. Then, system~\eqref{G14} reduces to 
\beq \label{G18b}
\mathcal{P}^2 = 4 \lambda^2 \mathcal{Q}_x^2\,, \qquad \frac{\mathcal{P}_{uu}}{\mathcal{P}} + \frac{\mathcal{Q}_{uu}}{\mathcal{Q}}=0\,.
\eeq
The resulting gravitational waves in this case all have metrics of the form 
\beq \label{G18c}
ds^2 = -dt^2 + dz^2  + 4\,\lambda^2\,\mathcal{Q}_x^2\, dx^2 + \mathcal{Q}^2\, dy^2\,,
\eeq
which turn out to represent \emph{plane} waves, since on the wave front $u$ must be replaced by a constant $u_0$, say, such that $\mathcal{Q}_x(x, u_0)$ and $\mathcal{Q}(x, u_0)$ are then only functions of $x$; hence, the resulting metric on the wave front,
\begin{equation}\label{G18d}
 d\sigma^2 = 4\, \lambda^2\, d\mathcal{Q}^2 + \mathcal{Q}^2\,dy^2\,,
\end{equation}
is flat. Moreover, with $k = \partial_v = \partial_t + \partial_z$, it is simple to see that $\mathbb{T}_{\mu \nu} = k_{[\mu , \nu]} = 0$, so that the rays are twistfree, as expected. 

System~\eqref{G18b} can be further reduced to 
\beq \label{G18e}
\mathcal{P} = \pm 2 \lambda \mathcal{Q}_x\,, \qquad (\mathcal{Q}\,\mathcal{Q}_{uu})_x = 0\,.
\eeq
Exact solutions of $\mathcal{Q}\,\mathcal{Q}_{uu} = \varpi(u)$, i.e. solutions that can be explicitly expressed in terms of familiar functions, are unknown; however, it is possible to solve this equation when $\varpi (u) = \kappa$, where $\kappa$ is a constant. In this case, 
$\mathcal{Q}_{uu} = \kappa / \mathcal{Q}$ can be integrated once to yield
\beq \label{G18f}
\mathcal{Q}_u^2 =  2 \,\kappa\,\ln{\mathcal{Q}} + \mathcal{E}(x)\,, 
\eeq
where $\mathcal{E}(x)$ is a certain energy function. Assuming $\mathcal{Q}$ is a monotonically increasing function of $u$, we can write
\beq \label{G18g}
\int_{\mathcal{Q}(x, 0)}^{\mathcal{Q}(x, u)} \frac{dZ}{\sqrt{2\,\kappa\,\ln {Z} +\mathcal{E}(x)}}  =  u\,, 
\eeq
where the integral can be expressed in terms of the error function. These results contain an infinite class of solutions that represent \emph{plane} gravitational waves propagating along the $z$ direction.

\subsection{$A_x + 2\,C_x = 0$}

It follows from this condition that $A+2\,C$ is only a function of $u$. Moreover, with $A_x=-2\,C_x$, Eqs.~\eqref{G6} and~\eqref{G7} both reduce to $B_x+ 3\,C_x =  2\,C_{xx}/C_x$, while Eq.~\eqref{G8} implies that 
$B_u+3\,C_u = 2\,C_{xu}/C_x$. Thus, $B+3\,C -2\,\ln{C_x}$ is simply a constant. Therefore, the field equations reduce in this case to 
\begin{equation} \label{G19}
\mathcal{A}= \frac{\alpha(u)}{\mathcal{C}^2}\,,   \qquad  \mathcal{B} = \lambda'^2\, \frac{\mathcal{C}_x^2}{\mathcal{C}^5}\,
\end{equation}
and Eq.~\eqref{G11}. Here, $\alpha(u)$ is an integration function and, as before, $\lambda' > 0$ is a constant. The resulting gravitational waves in this case all have metrics of the form
\beq \label{G19a}
ds^2 =\frac{\alpha(u)}{\mathcal{C}^2}\,( -dt^2 + dz^2 ) + \lambda'^2\,\frac{\mathcal{C}_x^2}{\mathcal{C}^5}\, dx^2 + \mathcal{C}\, dy^2\,,
\eeq
which turn out to represent \emph{nonplanar} waves with wave fronts that have \emph{negative} Gaussian curvature. Indeed, following the same line of argument as in Eqs.~\eqref{G18c} and~\eqref{G18d}, the wave front has a metric of the form
\begin{equation}\label{G19b}
 d\sigma^2 = \lambda'^2\,\frac{d\mathcal{C}^2}{\mathcal{C}^5} + \mathcal{C}\, dy^2\,,
\end{equation}
which has negative Gaussian curvature $K_G = - \mathcal{C}^3(x, u_0)/ \lambda'^2$, as discussed in Appendix B. Moreover, the twist  tensor in this case has nonzero components $\mathbb{T}_{01} = -\mathbb{T}_{10} = \mathbb{T}_{13} = -\mathbb{T}_{31} =\alpha(u)\,\mathcal{C}_x/\mathcal{C}^3$.

Let us now return to system~\eqref{G19} and note that in the corresponding general TGW metric~\eqref{G19a}, it is possible to absorb $\alpha(u)$ in the redefinition of the retarded null coordinate. Thus we can, in effect, set $\alpha(u)= 1$ in our system. Our general TGW metric  turns out to be of type \emph{II} in the Petrov classification. 

Defining $\mathcal{P}$ and $\mathcal{Q}$ as in Eq.~\eqref{G18a},  we find
\begin{equation}\label{G20}
\mathcal{A} = \mathcal{Q}^{-4}\,, \qquad  \mathcal{P} = \pm 2 \lambda' \frac{\mathcal{Q}_x}{\mathcal{Q}^4}\,,
\end{equation}
so that Eq.~\eqref{G11} takes the form
\beq \label{G21}
 \frac{\mathcal{P}_{uu}}{\mathcal{P}} + \frac{\mathcal{Q}_{uu}}{\mathcal{Q}}= - 4\,\frac{\mathcal{Q}_u}{\mathcal{Q}}\, \left(\frac{\mathcal{P}_u}{\mathcal{P}}+\frac{\mathcal{Q}_u}{\mathcal{Q}}\right)\,.
\eeq
Using the fact that 
\beq \label{G22}
 \frac{\mathcal{P}_{u}}{\mathcal{P}} =  \frac{\mathcal{Q}_{xu}}{\mathcal{Q}_x} - 4\,\frac{\mathcal{Q}_u}{\mathcal{Q}}\,,
\eeq
we can reduce the problem to the following differential equation for $\mathcal{Q}$,
\beq \label{G23}
 \mathcal{Q}^2\,\mathcal{Q}_{xuu} - 4\, \mathcal{Q}\,\mathcal{Q}_u\,\mathcal{Q}_{xu}  -  (3\, \mathcal{Q}\, \mathcal{Q}_{uu} - 8\, \mathcal{Q}_u^2)\,\mathcal{Q}_x = 0\,.
\eeq

Equation~\eqref{G23} turns out to be equivalent to 
\beq \label{G24}
\left(\frac{\mathcal{Q}_{uu}}{\mathcal{Q}^3} -2\, \frac{\mathcal{Q}_u^2}{\mathcal{Q}^4}\right)_x = 0\,.
\eeq
Hence, we have 
\beq \label{G25}
\frac{\mathcal{Q}_{uu}}{\mathcal{Q}^3} -2\, \frac{\mathcal{Q}_u^2}{\mathcal{Q}^4} = - \upsilon(u)\,,
\eeq
where $\upsilon(u)$ is an integration function. Let us note that Eq.~\eqref{G25} can be written as
\beq \label{G26}
\left(\frac{\mathcal{Q}_{u}}{\mathcal{Q}^2}\right)_u = - \upsilon(u)\,\mathcal{Q}\,. 
\eeq
Let us define $\mathcal{S}$ such that 
\beq \label{G27}
\mathcal{S} := \frac{1}{\mathcal{Q}}\,; 
\eeq
then, Eq.~\eqref{G26}  takes the form
\beq \label{G28}
\mathcal{S}_{uu} = \upsilon(u)\,\mathcal{S}^{-1}\,. 
\eeq
It is remarkable that we encounter here in case (B) the same type of equation that we encountered in case (A). As before, exact solutions of Eq.~\eqref{G28}, namely, solutions expressible in terms of familiar functions, are not known, except when $\upsilon(u)$ is a constant $k$. 
With  $\upsilon(u) = k$, Eq.~\eqref{G28} can be integrated once to yield
\beq \label{G29}
\mathcal{S}_u^2 =  2 \,k\,\ln{\mathcal{S}} + E(x)\,, 
\eeq
where $E(x)$ is a new energy function. Assuming $\mathcal{S}$ is a monotonically increasing function of $u$, we can write
\beq \label{G30}
\int_{\mathcal{S}(x, 0)}^{\mathcal{S}(x, u)} \frac{dZ}{\sqrt{2\,k\,\ln {Z} +E(x)}}  =  u\,, 
\eeq
where the integral can be expressed in terms of the error function. We have thus demonstrated the existence of an infinite class of TGWs, namely, exact unidirectional nonplanar gravitational wave solutions of general relativity theory. 

Finally, as is evident from the tenor of our work, we simply attempt to develop a physical idea in this paper. On the other hand, we should mention that our solutions belong to the Kundt class~\cite{R1, Griffiths:2009dfa} and are probably all known in some other coordinate systems and within certain classification schemes; in this connection, see Ref.~\cite{Kinn} and sections 24.4.1 and 31.5.2 of Ref.~\cite{R1}.

\section{Simple TGW Solutions}

To look for manageable TGW solutions, we solve Eq.~\eqref{G26}  for $\upsilon = 0$.  In view of Eq.~\eqref{G28}, the general solution is given by
\beq \label{H1}
\mathcal{Q} = \frac{1}{u\,f(x) + h(x)}\,, 
\eeq
where $f(x)$ and $h(x)$ are integration functions and cannot both be zero functions simultaneously. 
The spacetime metric can be worked out using Eq.~\eqref{G20} and the result is
\begin{equation}\label{H2}
 ds^2 = (u\,f+h)^4 (- dt^2 + dz^2) + 4\,\lambda'^2\,(u\,f + h)^4 \left(u\, \frac{df}{dx} + \frac{dh}{dx}\right)^2 \,dx^2 + (u\,f + h)^{-2}\,dy^2\,.
\end{equation}
Taking advantage of the freedom in the choice of the $x$ coordinate in Eq.~\eqref{H2}, we have a simple TGW solution that depends upon only one arbitrary function of $x$. It turns out that solution~\eqref{H2} is of type \emph{D} in the Petrov classification. Of this class of solutions, we are interested in this section in three special cases where $f(x) = 0$, $h(x) = 0$ and, finally, $f(x) = 1/q$ and $h(x) = x/q$, where $q$ is a constant. Let us now examine these cases in turn.

\subsection{$f(x) = 0$, $\mathcal{Q} = 1/h(x)$}

The spacetime metric in this case takes the form
\begin{equation}\label{H3}
 ds^2 = - \chi^4\,(dt^2-dz^2) + 4\,\lambda'^2\,\chi^4\,d\chi^2 + \chi^{-2}\,dy^2\,,
\end{equation}
where we have replaced $h(x)$ by $\chi$. The result is a spacelike Kasner metric~\cite{R1}; indeed, with a new spacelike coordinate $x$, 
\begin{equation}\label{H4}
x := \frac{2\,\lambda'}{3}\, \chi^3\,,
\end{equation}
and constant rescalings of the spacetime coordinates, the metric takes the standard \emph{static} form~\cite{R2}
\begin{equation}\label{H5}
 ds^2 = - x^{2p_1} dt^2 +dx^2 + x^{2p_2} dy^2 + x^{2p_3} dz^2\,,
\end{equation}
with $p_1+p_2+p_3=p_1^2+p_2^2+p_3^2 =1$. In our case,  
\begin{equation}\label{H6}
 p_1=p_3 =\frac{2}{3}\,, \qquad p_2 = - \frac{1}{3}\,,
\end{equation}
so that we can write our solution in the convenient form
\begin{equation}\label{H7}
ds^2 =  x^{4/3}\,(- dt^2 + dz^2) + dx^2+ x^{-2/3}\,dy^2\,.
\end{equation}
This particular \emph{static} Kasner spacetime is of type \emph{D} in the Petrov classification and its three independent Killing vectors are $\partial_v = \partial_t +\partial_z$, $\partial_y$ and $\partial_z$; moreover, its homothetic vector field is given by $t\, \partial_t + 3\,x\, \partial_x\, + 4\, y\, \partial_y+ z\, \partial_z$.

\subsection{$h(x) = 0$, $\mathcal{Q} = 1/[u\,f(x)]$}

The spacetime metric in this case can be expressed as
\begin{equation}\label{H8}
 ds^2 = - u^4\,\chi^4\,du\,dv + 4\,\lambda'^2\,u^6\,\chi^4\,d\chi^2 + u^{-2}\,\chi^{-2}\,dy^2\,,
\end{equation}
where we have replaced $f(x)$ by $\chi$. Furthermore, designating $u^5/5$ as $u$ in metric~\eqref{H8}, using Eq.~\eqref{H4} and constant rescalings of the coordinates finally lead to 
\begin{equation}\label{H9}
ds^2 =  x^{4/3}\,( - dt^2 + dz^2) + u^{6/5}\,dx^2+ x^{-2/3}\,u^{-2/5}\,dy^2\,,
\end{equation}  
which is Harrison's TGW solution~\cite{Harrison:1959zz, DIRC}. Harrison's spacetime has independent Killing vectors $\partial_v$, $\partial_y$ and a homothetic vector field $5\, t\, \partial_t + 6\,x\, \partial_x\, + 12\, y\, \partial_y+ 5\,z\, \partial_z$.
The wave front in this case has a metric of the form $d\sigma^2 =  dx^2 + x^{-2/3}\,dy^2$
and a Gaussian curvature $K_G = -4/(9\,x^2)$.

\subsection{$f(x) = 1/q$, $h(x) = x/q$ and $\mathcal{Q} = q/(x+u)$}

The spacetime metric in this case is given by
\begin{equation}\label{H10}
 ds^2 = \frac{1}{q^2}\,(x+u)^4 (- dt^2 + dz^2) + \frac{4\, \lambda'^2}{q^6}\,(x+u)^4\, dx^2 + q^2\,(x+u)^{-2} \,dy^2\,. 
\end{equation}
Let us first rescale the $y$ coordinate by replacing $q\,y$ by $y$. Next, we define positive constants $\lambda_0$ and $\lambda$ via
\begin{equation}\label{H11}
 \lambda_0 := \frac{1}{q^2}\,, \qquad \lambda := \frac{2\,\lambda'}{|q|^3}\,. 
\end{equation}
In this way, the metric of the new TGW solution takes the final form
\begin{equation}\label{H12}
 ds^2 = \lambda_0\,(x+u)^4 (- dt^2 + dz^2) +  \lambda^2\,(x+u)^4\, dx^2 + (x+u)^{-2} \,dy^2\,. 
\end{equation}
This is of type \emph{D} in the Petrov classification and represents a nonplanar gravitational wave propagating in the $z$ direction. Its wave front has metric
\begin{equation}\label{H13}
 d\sigma^2 = \lambda^2\,(x+u_0)^4\, dx^2 + (x +u_0)^{-2}\,dy^2\,,
\end{equation}
where $u_0$ is a constant. The Gaussian curvature of this surface is negative as well and is given by (cf. Appendix B)
\begin{equation}\label{H14}
 K_G =  -\frac{4}{\lambda^2\,(x+u_0)^6}\,.
\end{equation}
This TGW solution is further discussed in the next section. 

Of the three cases we have discussed in this section, only the Harrison solution and the new solution represent propagating gravitational waves. The special Kasner solution is static and represents a limiting form of these solutions; that is, if we replace $u$ by a constant $u_0 \ne 0$ in the metric functions of the Harrison and new solutions, then the resulting metrics are equivalent to the special Kasner metric via constant rescalings of the spacetime coordinates.  Similarly, if we replace $x$ by a constant $x_0 \ne 0$  in the  
metric functions of the Harrison and new solutions, we recover in the same manner the metric of the
 \emph{plane} gravitational wave
\beq \label{H14a}
ds^2 = - dt^2 + u^{6/5} dx^2 +  u^{-2/5} dy^2 + dz^2\,,
\eeq
which is a special case of the type \emph{N} metrics discussed in Section V of Ref.~\cite{Bini:2017qnd} with $s_2 = 3/5$ and $s_3 = -1/5$. Thus our simple TGW solutions may be considered to be nonlinear superpositions of this special linearly polarized plane gravitational wave and the special static Kasner solution~\eqref{H7}.

A Ricci-flat solution in GR has four algebraically independent scalar polynomial curvature invariants  given by~\cite{R1}
\begin{equation}\label{H15}
I_1 = R_{\mu \nu \rho \sigma}\,R^{\mu \nu \rho \sigma} - i R_{\mu \nu \rho \sigma}\,R^{*\,\mu \nu \rho \sigma}\,
\end{equation}  
and
\begin{equation}\label{H16}
I_2 = R_{\mu \nu \rho \sigma}\,R^{\rho \sigma \alpha \beta}\,R_{\alpha \beta}{}^{\mu \nu} + i R_{\mu \nu \rho \sigma}\,R^{\rho \sigma \alpha \beta}\,R^{*}{}_{\alpha \beta}{}^{\mu \nu}\,.
\end{equation}
The class of spacetimes under consideration here are all algebraically special of type \emph{D}; hence, $I_1^3 = 12\,I_2^2$. In the case of the general solution~\eqref{H2}, we find
\begin{equation}\label{H17}
I_1= \frac{12}{\lambda'{}^4  (u f +h)^{12}}\,, \qquad I_2=-\frac{12}{\lambda'{}^6  (u f +h)^{18}}\,,
\end{equation}
so that the timelike hypersurface $u\,f(x) + h(x) = 0$ is the curvature singularity of this spacetime.
More specifically, for the new solution~\eqref{H12}, we find     
\begin{equation}\label{H18}
I_1 = \frac{192}{\lambda^4\,(x+u)^{12}}\,, \qquad I_2 = -  \frac{768}{\lambda^6\,(x+u)^{18}}\,,
\end{equation}
so that  the timelike hypersurface $x+u=0$ is a curvature singularity in this case. For the Harrison solution~\eqref{H9}, we have
\begin{equation}\label{H19}
I_1 = \frac{64}{27\,x^4\,u^{12/5}}\,, \qquad I_2 = -  \frac{256}{243\,x^6\,u^{18/5}}\,,
\end{equation}
so that the timelike hypersurface $w:=x^{1/3}\,u^{1/5} = 0$ is the curvature singularity of the Harrison solution. Finally, in the limiting case of the special Kasner spacetime~\eqref{H7}, the curvature invariants are 
given by $I_1 = 64/(27\,x^4)$ and $I_2 = -  256/(243\,x^6)$;
hence, the timelike hypersurface $x=0$ is the curvature singularity of the spacelike Kasner metric. 

The twisted gravitational wave solutions have nonzero twist tensor. To see this explicitly, we note that for the new solution~\eqref{H12}, the twist tensor can be expressed as
\beq \label{H20}
(\mathbb{T}_{\mu \nu})
=\Phi   
\begin{pmatrix}
0& -1& 0& 0\cr
1& 0& 0& -1\cr
0& 0& 0& 0\cr
0& 1& 0& 0 \cr
\end{pmatrix}\,,\qquad \Phi = 2\lambda_0 \,(x+u)^3\,, 
\eeq
while for the Harrison solution the only difference is that  $\Phi = \frac23 \, x^{1/3}$.

\section{Timelike and null geodesics of the new TGW solution~\eqref{H12}}

To gain insight into the nature of a twisted gravitational wave  spacetime, it is useful to study the motion of free test particles in such a gravitational field. For this purpose, we choose the new TGW solution~\eqref{H12} and define $W$ via 
\beq \label{T1}
W = x + u\,,
\eeq
so that the metric  of the new TGW can be written as
\begin{equation}\label{T2}
 ds^2 = -\lambda_0\,W^4 (dt^2-dz^2) +  \lambda^2\,W^4\, dx^2 + W^{-2} \,dy^2\,. 
\end{equation}
This spacetime has one null and two spacelike Killing vector fields given by
\begin{equation}\label{T3}
 \partial_v =  \partial_t +  \partial_z\,, \qquad  \partial_x+  \partial_z\,, \qquad  \partial_y\,,
\end{equation}
respectively, and a homothetic vector field
\begin{equation}\label{T4}
 t\, \partial_t + x\, \partial_x\, + 4\, y\, \partial_y+ z\, \partial_z\,.
\end{equation}
We note that $\sqrt{-g} = \lambda_0\,\lambda\,W^5$, which vanishes at the curvature singularity $W=0$. 

The motion of free test particles and rays of radiation in this TGW spacetime involves three constants that can be obtained from the projection of the 4-velocity vector, $\dot{x}^\mu = dx^\mu/d\eta$, of the test particle on the null and spacelike Killing vector fields. Here, $\eta$ is either the proper time along a timelike geodesic or an affine parameter along the path of the null geodesic ray. Thus we have 
\begin{equation}\label{T5}
\frac{du}{d\eta} = \dot{t}-\dot{z} = \frac{C_v}{W^4}\,, \qquad \lambda^2\, \dot{x} + \lambda_0\, \dot{z} = \frac{C_0}{W^4}\,, \qquad \dot{y} = C_y\,W^2\,, 
\end{equation}
where $C_v$, $C_0$ and $C_y$ are constants of the motion. Let us note here that a special solution of the null geodesic equation corresponds to the Killing vector field $ \partial_v$ with $C_v = 0$, $C_y = 0$ and constant $W$ along the rays that indicate the propagation of the background TGW. For other null geodesics, we require that $C_v \ne 0$. Likewise, we are interested in future-directed timelike  geodesics; hence, we assume that $C_v > 0$, so that for constant $z$ we have $dt/d\eta > 0$. Moreover, $g_{\mu\nu}\,\dot{x}^\mu\, \dot{x}^\nu = - \epsilon$, where $\epsilon = 1$ or $\epsilon = 0$, depending upon whether the geodesic is timelike or null, respectively; that is, 
\begin{equation}\label{T6}
 \lambda_0\,W^4 (-\dot{t}^2+\dot{z}^2) +  \lambda^2\,W^4\, \dot{x}^2 + W^{-2} \,\dot{y}^2 = - \epsilon\,, \qquad \epsilon =1,0\,. 
\end{equation}

The geodesic equations of motion can be simply obtained from a Lagrangian of the form $(ds/d\eta)^2$. In this way, we find that the geodesic equation for the $x$ coordinate takes the form
\begin{equation}\label{T7}
\frac{d}{d\eta}[\lambda^2\,W^4 \dot{x}] = \frac{2\epsilon}{W} - 3 C_y^2\,W\,, 
\end{equation}
where Eq.~\eqref{T6} has been employed. Using $W^4 du/d\eta = C_v >0$, we can express Eq.~\eqref{T7} as
\begin{equation}\label{T8}
\frac{d^2x}{du^2} + 2\,\zeta \,W^3 + 3\,\zeta' \,W^5 = 0\,,
\end{equation}
where $\zeta$ and $\zeta'$ are constants given by
\begin{equation}\label{T9}
\zeta = \frac{\epsilon}{\lambda^2\,C_v^2}\,, \qquad  \zeta' = \frac{C_y^2}{\lambda^2\,C_v^2}\,.
\end{equation}
We recall that $W = x + u$ by definition; therefore, Eq.~\eqref{T8} can be integrated once and the result is
\begin{equation}\label{T10}
\left(\frac{dW}{du}\right)^2 + \zeta\, W^4 +\zeta'\,W^6 = \mathbb{E}\,,
\end{equation}
where $\mathbb{E}$ must therefore be positive and is given by
\begin{equation}\label{T11}
\mathbb{E} = 1 + \frac{\lambda_0\,C_v + 2\,C_0}{\lambda^2\,C_v}\,,
\end{equation}
which follows from Eqs.~\eqref{T5}--\eqref{T10}. Equation~\eqref{T10} bears a remarkable resemblance to Eq.~(95) of Section VI of Ref.~\cite{Bini:2017qnd}, which involved  the geodesic equation for the Harrison TGW solution. As in~\cite{Bini:2017qnd}, one can interpret Eq.~\eqref{T10} via a one-dimensional motion of a classical particle with positive energy $\mathbb{E}$ in a simple positive symmetric effective potential well of the same structure as in the case of Harrison's TGW solution. The motion is periodic with turning points at $\pm W_0$, where $W_0 >0$ and $\pm W_0$ are the only real solutions of Eq.~\eqref{T10} with $dW/du = 0$. We recall that $W = 0$ is the location of the spacetime singularity; therefore, the geodesic motion can start from $-W_0$ or $W_0$ and end  up at the curvature singularity. 

\subsection{Oblique Cosmic Jet}

Plane gravitational wave spacetimes have parallel rays and are of type \emph{N} in the Petrov classification. This means that the four principal null directions of the Weyl tensor coincide and are all parallel to the direction of propagation of the plane wave and hence perpendicular to the uniform wave front. The timelike geodesics of these spacetimes have the peculiar property that they all asymptotically line up 
parallel to the direction of motion of the wave and their Lorentz factors approach infinity. This \emph{cosmic jet} property was first demonstrated in Ref.~\cite{Bini:2014esa} and further elaborated in Ref.~\cite{Bini:2017qnd}. In the case of Harrison's nonplanar GW, the nonuniformity of the wave front led to an \emph{oblique} cosmic jet, namely, the direction of the cosmic jet deviated from the direction of propagation of the wave. On the other hand, Harrison's TGW solution, just like the new TGW solution under consideration in this section, is of type \emph{D} in the Petrov classification, which means that the four principal null directions in this case indicate only two directions (each with multiplicity 2): one along the direction of wave propagation and another along some oblique direction. 

It is possible that oblique cosmic jets occur for all TGW spacetimes.  Therefore, it would be interesting to see if the same result holds for the new TGW solution as well. To this end, let us refer the motion of timelike (and null) geodesics to fiducial observers that are at rest in this spacetime. The natural tetrad frame of these static observers is given by
\begin{eqnarray}\label{T12}
e_{\hat 0} = \lambda_0^{-1/2}W^{-2}\,\partial_t\,, \qquad
e_{\hat 1} = \lambda^{-1} W^{-2}\,\partial_x\,,\qquad
e_{\hat 2} =  W\,\partial_y\,, \qquad
e_{\hat 3} = \lambda_0^{-1/2}W^{-2}\, \partial_z\,.
\end{eqnarray}
Projecting $\dot{x}^\mu = (\dot t, \dot x, \dot y, \dot z)$ on $e^{\mu}{}_{\hat \alpha}$ results in $\dot{x}^{\hat \alpha} = \Gamma (1, V_x, V_y, V_z)$, where
\begin{equation}\label{T13}
\Gamma = \sqrt{\lambda_0}\,W^{2}\, \dot t\,, \qquad V_x = \frac{\lambda}{\sqrt{\lambda_0}}\,\frac{\dot x}{\dot t}\,, \qquad V_y = \frac{C_y}{\sqrt{\lambda_0}\,W\,\dot{t}}\,, \qquad  V_z = \frac{\dot z}{\dot t}\,.
\end{equation}
It follows from Eqs.~\eqref{T5}--\eqref{T10} that 
\begin{equation}\label{T14}
\dot{t} = \frac{\epsilon+C_y^2\,W^2}{2\lambda_0 C_v} + \frac{C_v}{2W^4} \left [\frac{\lambda^2}{\lambda_0}\,\left(\frac{dx}{du}\right)^2 + 1\right]\,,
\end{equation}
\begin{equation}\label{T15}
\dot{x} = \frac{C_v}{W^4}\,\frac{dx}{du}\,,
\end{equation}
\begin{equation}\label{T16}
\dot{y} = C_y\,W^2\,
\end{equation}
and
\begin{equation}\label{T17}
\dot{z} = \frac{\epsilon+C_y^2\,W^2}{2\lambda_0 C_v} + \frac{C_v}{2W^4} \left [\frac{\lambda^2}{\lambda_0}\,\left(\frac{dx}{du}\right)^2 - 1\right]\,.
\end{equation}
Moreover, as $W \to 0$, $(dW/du)^2 \to \mathbb{E}$. 
As free test particles approach the spacetime singularity at $W=0$, we find that for $W \to 0$, a cosmic jet develops with $\Gamma \to \infty$ and
\begin{equation}\label{T18}
(V_x, V_y, V_z) \to  (\sin \Theta_{\pm}, 0, \cos \Theta_{\pm})\,, 
\end{equation}
where 
\begin{equation}\label{T19}
\cot \left(\frac{\Theta_{\pm}}{2}\right) = \frac{\lambda}{\sqrt{\lambda_0}}\,\left(\pm\sqrt{\mathbb{E}} - 1\right)\,. 
\end{equation}
In this equation, the upper (lower) sign indicates that the singularity at $W=0$ is approached from the turning point at $-W_0$ ($W_0$).
The oblique character of the cosmic jet in this case is in complete correspondence with Harrison's TGW. 

Are there observable differences between plane and twisted gravitational waves? To investigate this issue, we can compare and contrast the influence of these waves on the propagation of fields and on congruences of massive test particles. The case of a massless scalar field is treated in Appendix C. Tidal effects of TGWs are studied in the next section.

\section{Jacobi Equation}

Imagine the world line of an arbitrary reference observer that is static in the spacetime under consideration. Let $\tau$ be the proper time and $\lambda^{\mu}{}_{\hat \alpha}(\tau)$ be a Fermi-Walker transported tetrad along this world line, where at each event $\bar{x}^\mu(\tau)$ we imagine all spacelike geodesic curves that issue perpendicularly from this event and generate a local hypersurface. We assume that $x^\mu$ is an event on this hypersurface that can be connected to $\bar{x}^\mu(\tau)$ via a \emph{unique} spacelike geodesic of proper length $\varsigma$. We assign Fermi coordinates $X^{\hat \mu} = (T, X^{\hat i})$  to event $x^\mu$, where
\begin{equation}\label{J1}
T := \tau\,, \qquad X^{\hat i} := \varsigma\, \xi^\mu(\tau)\, \lambda_{\mu}{}^{\hat i}(\tau)\,.
\end{equation}
Here, $\xi^\mu(\tau)$ is the unit spacelike vector at $\bar{x}^\mu(\tau)$ that is tangent to the unique geodesic connecting $\bar{x}^\mu(\tau)$ to $x^\mu$,
so that $\xi^\mu(\tau)\,\lambda_{\mu}{}^{\hat 0}(\tau) = 0$. Thus along the reference world line, $\xi^\mu(\tau)\, \lambda_{\mu}{}^{\hat i}(\tau)$, for $i = 1, 2, 3$, are the corresponding direction cosines at proper time $\tau$.  In the Fermi coordinate system, the reference observer is permanently fixed at the spatial origin ($\mathbf{X} = 0$). We are interested in the equation of motion of a neighboring free test particle relative to the reference observer in the Fermi coordinate system. Neglecting the relative velocity, the reduced geodesic equation can be expressed as
\begin{equation}\label{J2}
\frac{d^2X^{\hat i}}{dT^2}+A^{\hat i}+(R_{\hat 0 \hat i \hat 0 \hat j}+A_{\hat i}\,A_{\hat j})X^{\hat j} = 0\,,
\end{equation}
where $A^{\hat i}$ is the measured 4-acceleration of the fiducial static observer and $R_{\hat 0 \hat i \hat 0 \hat j}$ are the corresponding components of the tidal matrix, namely, 
\begin{equation}\label{J3}
A^{\hat i}(T) = \frac{D\lambda^{\mu}{}_{\hat 0}}{d\tau}\,\lambda_{\mu}{}^{\hat i}\,, \qquad R_{\hat \alpha\hat \beta \hat \gamma \hat \delta}(T) := R_{\mu \nu \rho \sigma}\,\lambda^{\mu}{}_{\hat \alpha}\,
\lambda^{\nu}{}_{\hat \beta}\,\lambda^{\rho}{}_{\hat \gamma}\,\lambda^{\sigma}{}_{\hat \delta}\,.
\end{equation}   
For background material on equations of motion of free test particles in Fermi coordinates, we refer to our recent paper~\cite{Bini:2017uax}; further material is contained in~\cite{mas77, CM3}   and the references cited therein. The Fermi coordinate system is generally admissible in a cylindrical region of radius $|\mathbf{X}| \sim \mathcal{R}$ in the spacetime domain around the reference world line, where
$\mathcal{R}$ is a certain minimal radius of curvature. 

Let us first consider the simple case of motion in the special linearly polarized \emph{plane} gravitational wave given by metric~\eqref{H14a}.  In this case, the  curvature singularity of spacetime occurs at $u = 0$.  The static observers in this spacetime follow geodesics and have a natural tetrad frame that is parallel transported along their geodesic world lines, namely,
\begin{eqnarray}\label{J5}
\lambda_{\hat 0} = \partial_t\,, \qquad
\lambda_{\hat 1} = u^{-3/5}\,\partial_x\,,\qquad
\lambda_{\hat 2} =  u^{1/5}\,\partial_y\,, \qquad
\lambda_{\hat 3} = \partial_z\,.
\end{eqnarray} 
Let us choose a static observer with world line $(t, x, y, z) = (\tau, 0, 0, 0)$, along which we construct a Fermi normal coordinate system. Using the results of Appendix B of  Ref.~\cite{Bini:2017qnd} for the curvature of the plane wave, the equations of motion of nearby free test particles in Fermi coordinates take the form
\begin{equation}\label{J6}
\frac{d^2X^{\hat 1}}{dT^2}+ \frac{6}{25\,T^2}\, X^{\hat 1} = 0\,,
\end{equation}
\begin{equation}\label{J7}
\frac{d^2X^{\hat 2}}{dT^2} - \frac{6}{25\,T^2}\, X^{\hat 2} = 0\,
\end{equation}
and $d^2 X^{\hat 3}/dT^2 = 0$. With the initial conditions that at $T = 1$, $(X^{\hat 1}, X^{\hat 2}, X^{\hat 3}) = (X_0, Y_0, Z_0)$ and $dX^{\hat i}/dT = 0$ for $i = 1, 2, 3$, we find
\begin{equation}\label{J8}
X^{\hat 1} = X_0 \,(-2 \,T^{3/5} + 3 \,T^{2/5})\,, \qquad X^{\hat 2} = \frac{1}{7}\,Y_0 \,(T^{6/5} + 6\, T^{-1/5})\,, \qquad X^{\hat 3} = Z_0\,.
\end{equation}
The plane wave in this case has the character of the plus ($\oplus$) polarization and for $T : 1 \to 0$, the particles move in the $(X^{\hat 1}, X^{\hat 2})$ plane such that at the curvature singularity $T = 0$, $(X^{\hat 1}, X^{\hat 2}, X^{\hat 3}) = (0, \infty, Z_0)$. Let us note an important limitation of our result here and in the rest of this section: Though we seek solutions of the Jacobi equation, the results are only valid so long as $|\mathbf{X}|$ is sufficiently small in accordance with the admissibility of Fermi normal coordinates. Thus in Eq.~\eqref{J8}, for instance, assuming that initially $|\mathbf{X}_0|$
is sufficiently small, the motion in $X^{\hat 2}$ is valid for only a very short time interval.

\subsection{Jacobi Equation for Harrison's TGW}

Next, we consider Harrison's twisted gravitational wave with metric
\beq \label{J9}
ds^2 = - x^{4/3}\,dt^2 + u^{6/5} dx^2 +  x^{-2/3}\,u^{-2/5} dy^2 + x^{4/3}\,dz^2\,,
\eeq
where we have set $\lambda = 1$ with no loss in generality. This spacetime has curvature singularities at $x = 0$ and $u = 0$. The static observers in this spacetime have a natural tetrad frame given by
\begin{eqnarray}\label{J10}
\lambda_{\hat 0} = x^{-2/3}\,\partial_t\,, \qquad
\lambda_{\hat 1} = u^{-3/5}\,\partial_x\,,\qquad
\lambda_{\hat 2} =  x^{1/3}\,u^{1/5}\,\partial_y\,, \qquad
\lambda_{\hat 3} = x^{-2/3}\,\partial_z\,.
\end{eqnarray}
The world line of a static observer in Harrison spacetime is accelerated, namely,  
\beq \label{J10a}
A=\nabla_{\lambda_{\hat 0}}\lambda_{\hat 0}=\frac{2}{3\,x\,u^{6/5}}\,\partial_x\,.
\eeq
Imagine that  $\mathbb{S}^\mu$ is a vector that is Fermi--Walker transported along  $\lambda^{\mu}{}_{\hat 0}$; then,  
\begin{equation}\label{J11}
\frac{d\mathbb{S}^\mu}{d\tau}+\Gamma^{\mu}_{\alpha \beta}\, \lambda^{\alpha}{}_{\hat 0}\,\mathbb{S}^{\beta}= (A\cdot \mathbb{S})\,\lambda^{\mu}{}_{\hat 0}-(\lambda_{\hat 0} \cdot \mathbb{S})\,A^\mu\,.
\end{equation}
It is straightforward to check that tetrad frame~\eqref{J10} is Fermi--Walker propagated along $\lambda_{\hat 0}$. Let us now consider the frame components of the 4-acceleration and of the relevant components of the electric part of the Riemann tensor. We find
\beq \label{J12}
A=\frac{2}{3\,x\,u^{3/5}}\, \lambda_{\hat 1}
\eeq
and the symmetric and traceless tidal matrix $\mathcal{K}$, 
\beq \label{J13}
\mathcal{K}_{\hat i \hat j} = R_{\hat 0\hat i \hat 0 \hat j}\,,
\eeq
has nonzero components,
\begin{eqnarray} \label{J14}
\mathcal{K}_{\hat1\hat1}&=&  -\frac{2}{9}\, \frac{1}{x^2\,u^{6/5}} +\frac{6}{25}\,\frac{1}{x^{4/3}\,u^2}\,, \nonumber\\
\mathcal{K}_{\hat 2\hat 2}&=&  -\frac{2}{9}\, \frac{1}{x^2\,u^{6/5}} -\frac{6}{25}\,\frac{1}{x^{4/3}\,u^2}\,,  \nonumber\\
\mathcal{K}_{\hat 3 \hat 3}&=&  \frac{4}{9\, x^2\,u^{6/5}}\,, \nonumber\\
\mathcal{K}_{\hat 1 \hat 3} &=& \mathcal{K}_{\hat 3 \hat 1} = \frac{2}{5\, x^{5/3}\, u^{8/5}}\,.
\end{eqnarray}

Let us now consider a static observer located at $(x_0, y_0, z_0)$, where $x_0 \ne 0$. Then, for this observer, $t = t_0 +\tau/x_0^{2/3}$ and $u = u_0 +\tau/x_0^{2/3}$, where $u_0 := t_0 -z_0$. We establish a Fermi normal coordinate system in the neighborhood of this observer. The equation of motion of a nearby free test particle in the Fermi system is given by Eq.~\eqref{J2}, where $A_{\hat i}$ and $\mathcal{K}_{\hat i \hat j}$ can be obtained from Eqs.~\eqref{J13} and~\eqref{J14} with 
\beq \label{J15}
 x = x_0\,, \qquad u = u_0 + \frac{T}{x_0^{2/3}}\,.
\eeq.

We assume for the sake of simplicity that $x_0 = 1$ and $u_0 = 0$. Equation~\eqref{J2} can then be written as
\begin{equation}\label{J16}
\frac{d^2X^{\hat 1}}{dT^2}+\frac{2}{3}\,\frac{1}{T^{3/5}} +\left(\frac{2}{9}\,\frac{1}{T^{6/5}} + \frac{6}{25}\,\frac{1}{T^2}\right)\, X^{\hat 1} + \frac{2}{5}\,\frac{1}{T^{8/5}} \, X^{\hat 3} = 0\,,
\end{equation}
\begin{equation}\label{J17}
\frac{d^2X^{\hat 2}}{dT^2} - \left(\frac{2}{9}\,\frac{1}{T^{6/5}} + \frac{6}{25}\,\frac{1}{T^2}\right)\, X^{\hat 2} = 0\,
\end{equation}
and
\begin{equation}\label{J18}
\frac{d^2X^{\hat 3}}{dT^2} + \frac{2}{5}\,\frac{1}{T^{8/5}}\, X^{\hat 1} + \frac{4}{9}\,\frac{1}{T^{6/5}}\, X^{\hat 3} = 0\,,
\end{equation}
which can be integrated from $T = 1$ to the curvature singularity at $T = 0$ with the boundary conditions that  at $T = 1$, $(X^{\hat 1}, X^{\hat 2}, X^{\hat 3}) = (X_0, Y_0, Z_0)$ and $dX^{\hat i}/dT = 0$ for $i = 1, 2, 3$. See, for example, Figure 1.

%%%%%%%%%%%%%%%%
\begin{figure}
\includegraphics[scale=0.3]{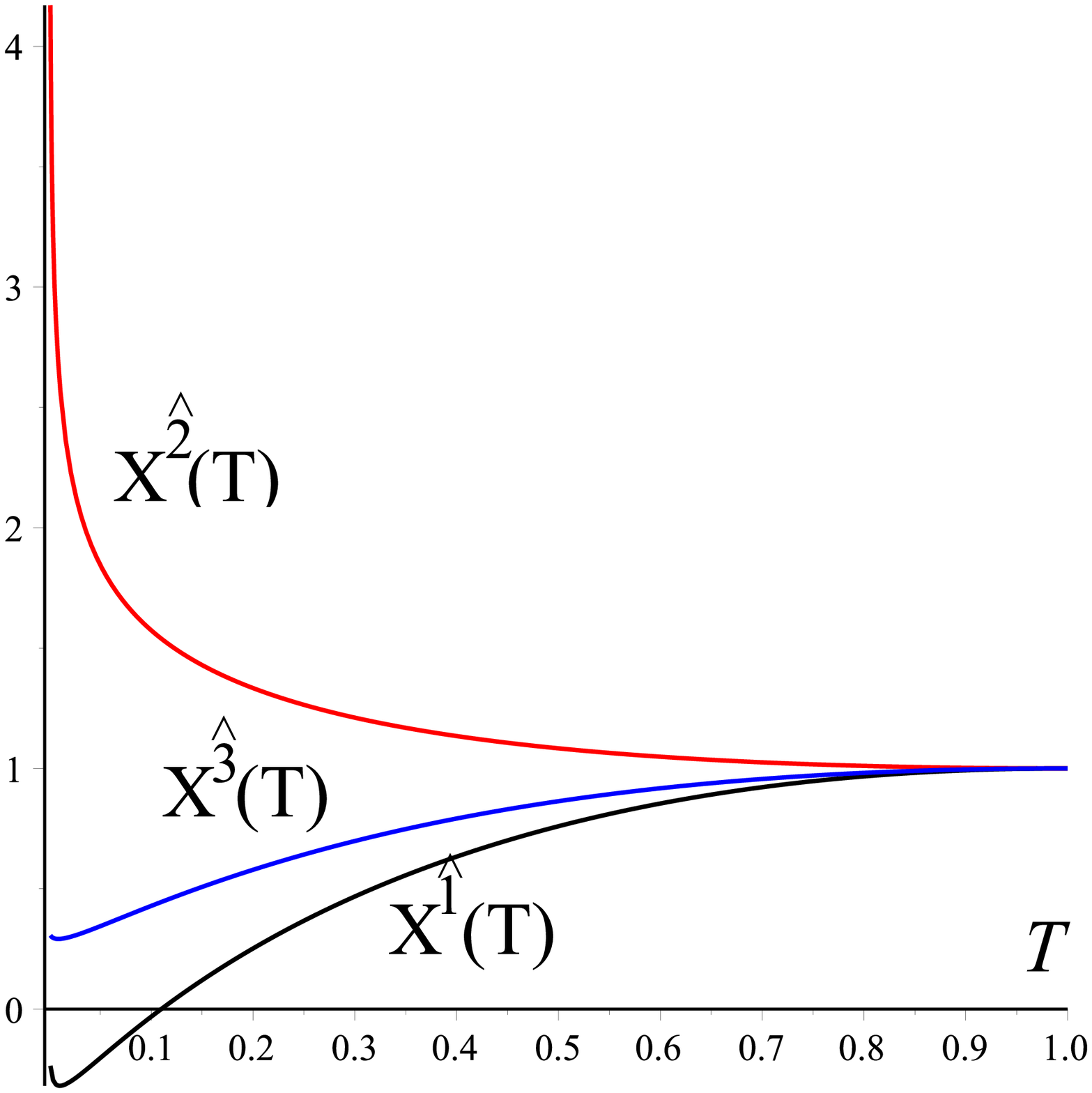}
\includegraphics[scale=0.3]{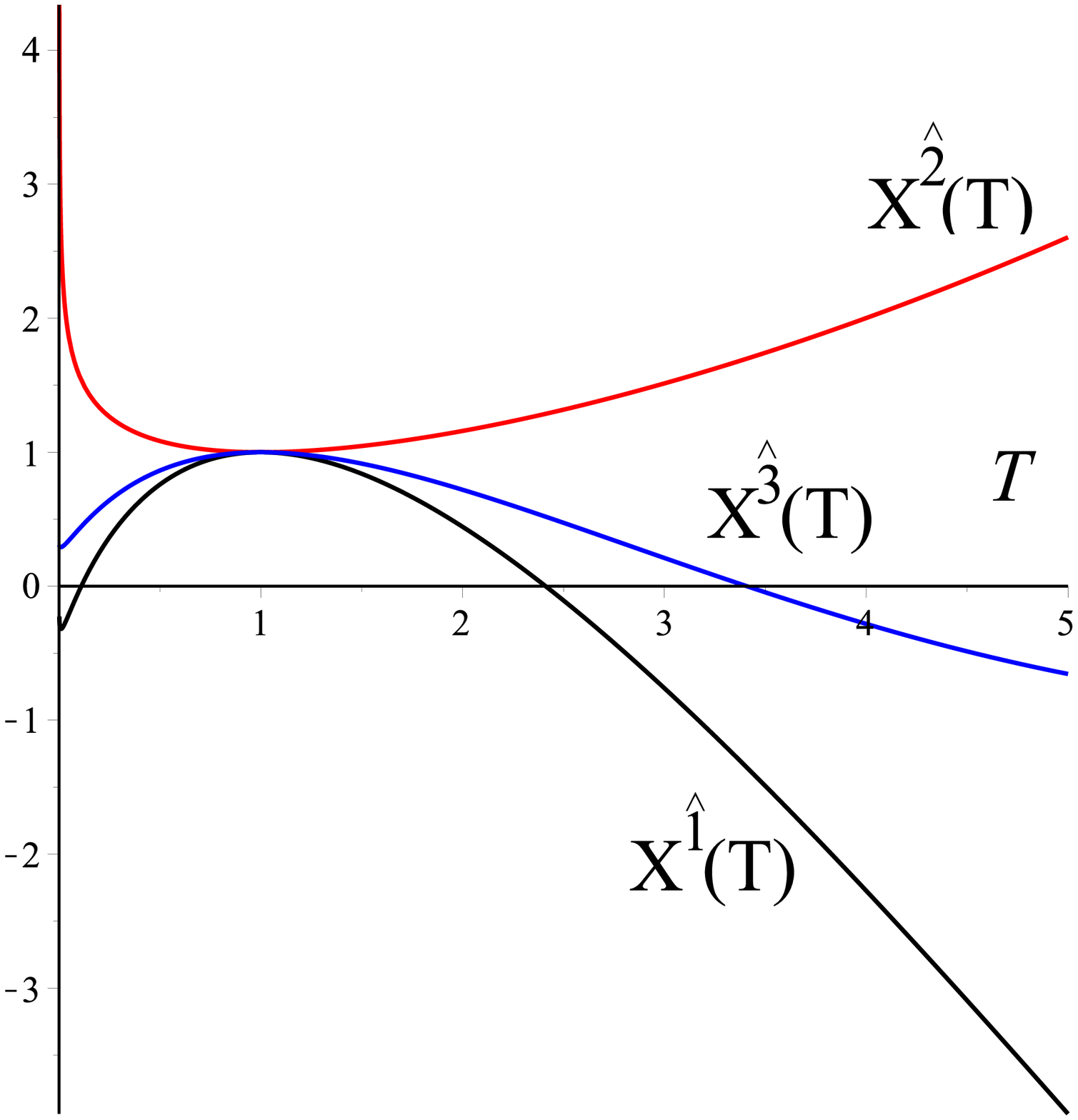}
\caption{\label{fig1a,fig1b} Numerical integration of  Eqs.~\eqref{J16}--\eqref{J18}. Initial conditions at $T = 1$ are chosen such that $X^{\hat i}=1$ and $dX^{\hat i}/dT=0$. Closer inspection reveals that $X^{\hat 1}$ indeed vanishes at $T = 0$, cf. Appendix D.}
\end{figure}
%%%%%%%%%%%%%%%%

It is the variation of $X^{\hat 3}$ with Fermi time $T$ that is the distinguishing feature of the TGW. To see this, let us first note that in system~\eqref{J16}--\eqref{J18}, the equation for $X^{\hat 2}$ decouples and can be treated separately. In fact, let 
\begin{equation}\label{J19}
X^{\hat 2} = T^{1/2}\, F_{\nu}(\vartheta)\,, \qquad  \vartheta = \frac{5}{3\sqrt{2}}\,T^{2/5}\,;
\end{equation}
then, Eq.~\eqref{J17} takes the form of the modified Bessel equation of order $\nu = 7/4$,
\begin{equation}\label{J20}
\vartheta^2\,\frac{d^2F_{\nu}}{d\vartheta^2} + \vartheta\,\frac{dF_{\nu}}{d\vartheta} - (\vartheta^2 + \nu^2) \,F_{\nu} = 0\,.
\end{equation}
Here,  $F_{\nu}(\vartheta)$ for $\nu = 7/4$ is a constant linear combination of $I_{7/4}(\vartheta)$ and 
$K_{7/4}(\vartheta)$. Moreover, as $T \to 0$, one can show from the properties of the modified Bessel functions of the first and second kind that $X^{\hat 2}$ diverges as $T^{-1/5}$, in agreement with Eq.~\eqref{J8} for the corresponding \emph{plane} gravitational wave. Next, for the $(X^{\hat 1}, X^{\hat 3})$ system, it proves convenient to introduce a new variable $S$,
\begin{equation}\label{J21}
S = - \ln{T}\,,
\end{equation}
so that as $T: 1 \to 0$, we have that $S: 0 \to \infty$. In terms of $S$, the $(X^{\hat 1}, X^{\hat 3})$ system can be expressed as
\begin{equation}\label{J22}
\frac{d^2X^{\hat 1}}{dS^2}+ \frac{dX^{\hat 1}}{dS} + \frac{2}{3}\,e^{-7S/2} +\left(\frac{2}{9}\,e^{-4S/5} + \frac{6}{25}\right)\, X^{\hat 1} + \frac{2}{5}\,e^{-2S/5} \, X^{\hat 3} = 0\,,
\end{equation}
\begin{equation}\label{J23}
\frac{d^2X^{\hat 3}}{dS^2} + \frac{dX^{\hat 3}}{dS} + \frac{2}{5}\,e^{-2S/5}\, X^{\hat 1} + \frac{4}{9}\,e^{-4S/5}\, X^{\hat 3} = 0\,.
\end{equation}
As proved in Appendix D,  all solutions of this system and their derivatives  are bounded for $0\le S< \infty$ and  $X^{\hat 1}$ converges to zero as $S\to \infty$. Numerical experiments reported in Appendix D suggest that $X^{\hat 3}$ also has a limit (which may not be zero) as $S\to \infty$.
Ignoring terms involving $e^{-2S/5}$, $e^{-4S/5}$ and $e^{-7S/2}$, system~\eqref{J22}--\eqref{J23} reduces to the corresponding system for the \emph{plane} GW~\eqref{H14a}. In fact, these terms vanish for $S \to \infty$. Thus the main difference between the TGW and the plane wave~\eqref{H14a} is that  in time  $X^{\hat 3}$  deviates from its value $Z_0$ at $S = 0$ in the case of TGW. The influence of the exact \emph{plane} wave~\eqref{H14a} on test particles that are initially at rest in the Fermi system has the same \emph{transverse} character as for linearized gravitational waves moving along the $z$ direction; however, the corresponding TGW has in addition a \emph{longitudinal} influence as well resulting in the $X^{\hat 3}$ component of the motion. This longitudinal feature of TGW and the associated temporal variation in $X^{\hat 3}$ may provide an observational signature for TGWs.

\subsection{Jacobi Equation for TGW~\eqref{H12}}

To confirm the longitudinal component of motion of free test particles induced by a TGW, it is interesting to study the Jacobi equation for the TGW solution~\eqref{H12}. In this case,
the static observers' adapted frame is given by Eq.~\eqref{T12}. It can be shown that this tetrad frame is Fermi-Walker transported  along $e_{\hat 0}$. In fact, $e_{\hat 0}$ has a nonzero 4-acceleration,
\beq \label{J24}
A=\nabla_{e_{\hat 0}}e_{\hat 0}=\frac{2}{W^5}\left(\frac{1}{\lambda^2}\partial_x -\frac{1}{\lambda_0}\partial_z\right)\,,
\eeq
where, as before, $W = x+u$. The projection of the 4-acceleration of static observers in this spacetime  on their tetrad frame field can be expressed as
\beq \label{J25}
A= \frac{2}{W^3} \left(\frac{1}{\lambda}  e_{\hat 1}  -\frac{1}{\lambda_0^{1/2}} e_{\hat 3}   \right)\,.
\eeq
Furthermore, tidal matrix~\eqref{J13} in this case has nonzero components given by
\begin{eqnarray}\label{J26}
\mathcal{K}_{\hat1\hat1}&=& \frac{2 (-\lambda_0+3\lambda^2)}{\lambda_0\lambda^2}\,W^{-6}\,,   \nonumber\\
\mathcal{K}_{\hat 2\hat 2}&=&  -\frac{2 (\lambda_0+3\lambda^2)}{\lambda_0\lambda^2}\,W^{-6}\,,  \nonumber\\
\mathcal{K}_{\hat 3 \hat 3}&=&  \frac{4}{\lambda^2}\,W^{-6}\,, \nonumber\\
\mathcal{K}_{\hat 1 \hat 3} &=& \mathcal{K}_{\hat 3 \hat 1} = \frac{6}{\lambda_0^{1/2}\lambda}\,W^{-6}\,.
\end{eqnarray}
Let us recall the fact that $W = 0$ corresponds to the curvature singularity in this TGW spacetime. 

We wish to establish a Fermi normal coordinate system in the neighborhood of a particular static observer; to this end, let us choose the observer located at $x^\mu = (t, x_0, y_0, z_0)$, where $x_0$, $y_0$ and $z_0$ are constants and $dt/d\tau = \lambda_0^{-1/2}W^{-2}$ in conformity with the expression for $e_{\hat 0}$ in Eq.~\eqref{T12}. Let us define $ \mathcal{W}$ to be the magnitude of $W = x + u$ along the world line of our reference static observer, namely, 
\beq \label{J26a}
 \mathcal{W} = x_0 + t - z_0\,,
\eeq
such that
\beq \label{J26b}
\frac{d\mathcal{W}}{d\tau} = \frac{1}{\sqrt{\lambda_0}\, \mathcal{W}^{2}}\,.
\eeq
Integrating this equation we find
\beq \label{J26c}
\sqrt{\lambda_0}\, \mathcal{W}^{3} = 3\,(\tau - \tau_0)\,, 
\eeq
where $\tau_0$ is an integration constant. To simplify matters, we choose $\tau_0 = 0$, so that
$\sqrt{\lambda_0}\, \mathcal{W}^{3} = 3\,\tau$
and the curvature singularity now occurs at $\tau = 0$ along the world line of our fiducial static observer. 

%%%%%%%%%%%%%%%%%%%%
\begin{figure}
\includegraphics[scale=0.3]{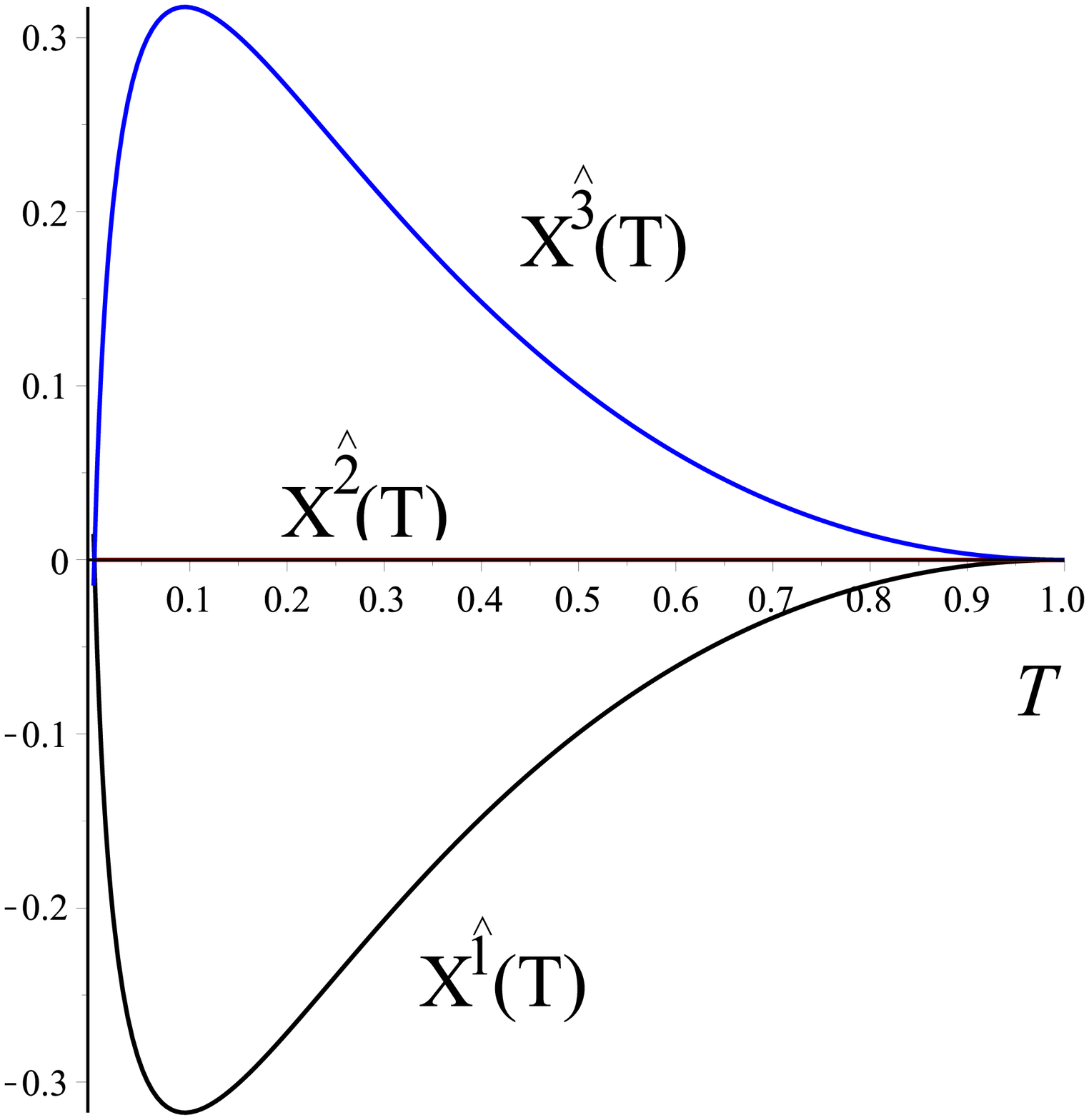}
\includegraphics[scale=0.3]{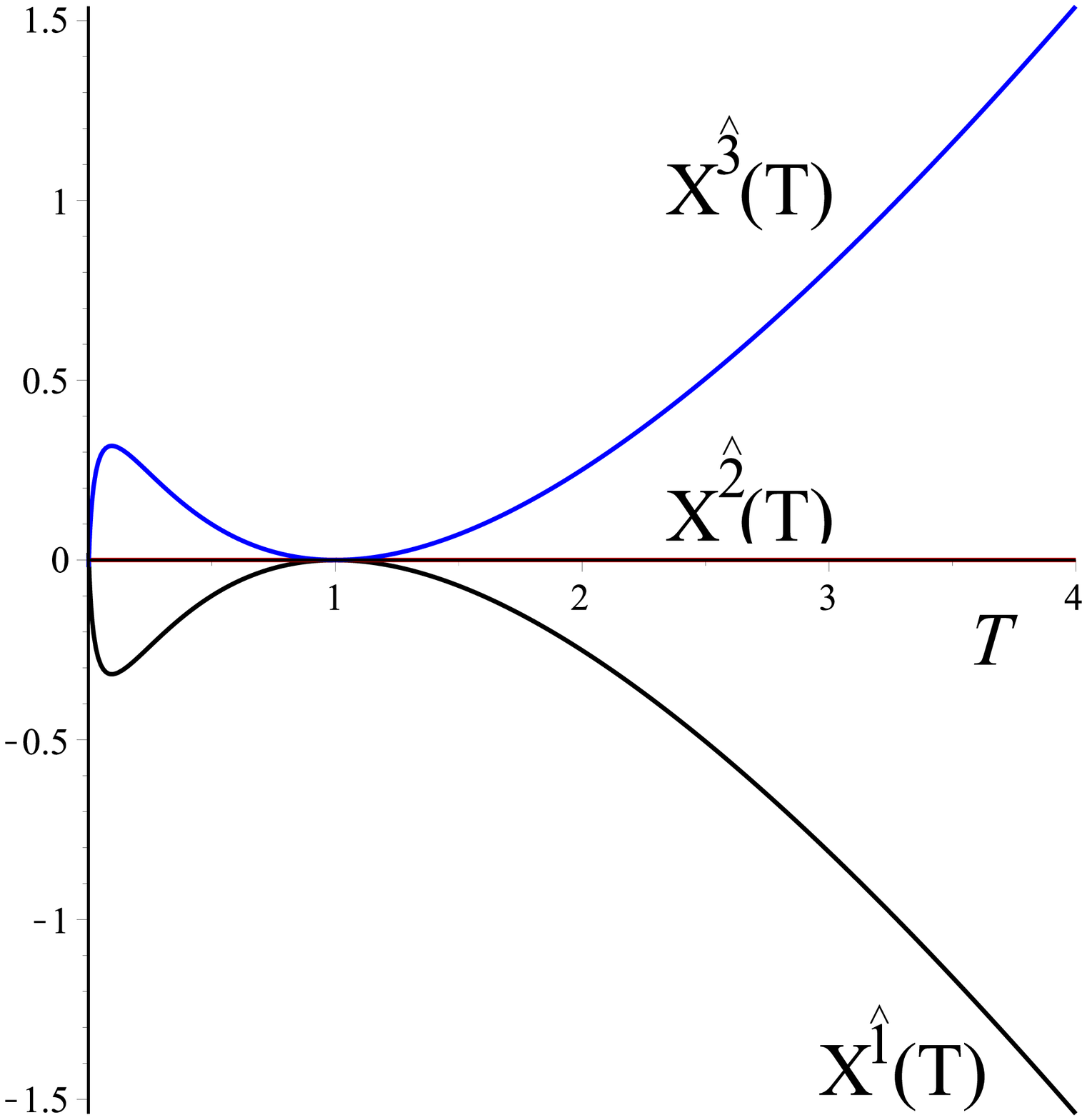}
\caption{\label{fig2a, fig2b} Numerical integration of  Eqs.~\eqref{J27}--\eqref{J29} with $\ell = 1$. Initial conditions are chosen so that $X^{\hat i} =0$, $dX^{\hat i}/dT=0$ at $T = 1$. With these conditions $X^{\hat 2}(T)=0$ identically.}
\end{figure}
%%%%%%%%%%%%%%%%%%%%

Next, we need to evaluate the 4-acceleration and the tidal matrix along the world line of the fiducial static observer by replacing $W$ with $\mathcal{W}$ in Eqs.~\eqref{J25} and~\eqref{J26}. The  reduced geodesic Eq.~\eqref{J2} then takes the form
\begin{equation}\label{J27}
\frac{d^2X^{\hat 1}}{dT^2}+\frac{2}{3}\,\frac{\ell}{T} + \frac{2}{9}\, \left(3+\ell^2\right)\,\frac{X^{\hat 1}}{T^2}\,  + \frac{2}{9}\,\ell\,\frac{X^{\hat 3}}{T^2} = 0\,,
\end{equation}
\begin{equation}\label{J28}
\frac{d^2X^{\hat 2}}{dT^2} -\frac{2}{9}\, \left(3+\ell^2\right)\,\frac{X^{\hat 2}}{T^2}  = 0\,
\end{equation}
and
\begin{equation}\label{J29}
\frac{d^2X^{\hat 3}}{dT^2} - \frac{2}{3}\,\frac{1}{T} +\frac{2}{9}\,\ell\,\frac{X^{\hat 1}}{T^2} + \frac{4}{9}\,(1+\ell^2)\frac{X^{\hat 3}}{T^2} = 0\,,
\end{equation}
where
\begin{equation}\label{J30}
\ell := \frac{\sqrt{\lambda_0}}{\lambda}\,.
\end{equation}

As before, the equation for $X^{\hat 2}$ decouples from the rest and its solution is given by
\begin{equation}\label{J31}
X^{\hat 2} = T^{1/2}\,(\check{b}_{+} \,T^{\check{a}} +\check{b}_{-} \, T^{-\check{a}})\,,
\end{equation}
where $\check{b}_{\pm}$ are integration constants and
\begin{equation}\label{J32}
\check{a} = \frac{1}{6}\,\sqrt{33+8\,\ell^2}\,.
\end{equation}

Let us introduce the independent variable $S = - \ln{T}$ into $(X^{\hat 1}, X^{\hat 3})$ system as before. The resulting system can be integrated using elementary methods and we find
\begin{equation}\label{J33}
X^{\hat 1} = \left[\mathbb{A}_1 \cos(\omega_1S + \varphi_1) + \mathbb{A}_2 \cos(\omega_2S + \varphi_2)\right]\,e^{-S/2} -\frac{3\,\ell}{\ell^2 + 2}\,e^{-S}\, 
\end{equation}
and 
\begin{equation}\label{J34}
X^{\hat 3} = \left[-\frac{1}{\ell}\,\mathbb{A}_1 \cos(\omega_1S + \varphi_1) + \ell \,\mathbb{A}_2 \cos(\omega_2S + \varphi_2)\right]\,e^{-S/2}+ \frac{3}{\ell^2 + 2}\,e^{-S}\,, 
\end{equation}
where
\begin{equation}\label{J35}
\omega_1 = \frac{1}{6}\,\sqrt{7+8\,\ell^2}\,, \qquad \omega_2 = \frac{1}{6}\,\sqrt{15+16\,\ell^2}\,
\end{equation}
and  $(\mathbb{A}_1$, $\varphi_1$, $\mathbb{A}_2$, $\varphi_2)$ are constants that can be determined from the initial conditions that we need to impose on $(X^{\hat 1}, dX^{\hat 1}/dS, X^{\hat 3}, dX^{\hat 3}/dS)$ at, say, $S = 0$. 

It is clear from Eqs.~\eqref{J33}--\eqref{J34} that $X^{\hat 1}$ and $X^{\hat 3}$ are bounded for $0\le S< \infty$ and vanish as $S\to \infty$. For the specific TGW under consideration here, the important quantity is the temporal variation of $X^{\hat 3}$ given by Eq.~\eqref{J34}. This is the longitudinal signature of the TGW that exists in addition to the transverse components characterized by the 
$(X^{\hat 1}, X^{\hat 2})$ components. 

Figure 2 depicts $X^{\hat 1}$ and $X^{\hat 3}$  given in Eqs.~\eqref{J33} and~\eqref{J34} for the special case of $\ell = 1$ and initial data  that  at $T = 1$, $X^{\hat i} = 0$ and $dX^{\hat i}/dT = 0$ for $i = 1, 3$. 

Finally, we should mention that the appearance of the longitudinal feature of TGWs via the Jacobi equation is not totally unexpected. In canonical tetrad frames, it is possible to use the Petrov classification to show that the behavior of the local free gravitational field---namely, the Weyl curvature tensor---within the framework of the geodesic deviation equation is  in general determined by the linear superposition of a transverse wave component, a longitudinal component and a Coulomb component~\cite{PZ, BP2, Podolsky:2012he}.

\section{DISCUSSION}

Linearized gravitational radiation can be represented as a linear superposition of monochromatic plane waves that are \emph{transverse} and exhibit plus $(\oplus)$ and cross $(\otimes)$ linear polarization states. In the nonlinear regime, exact solutions of GR that represent plane gravitational waves generally correspond to linearized plane waves propagating along a fixed direction in space.  On the other hand, exact gravitational waves in GR propagating along a fixed spatial direction can be \emph{nonplanar} as well. This paper is about such twisted gravitational waves (TGWs). We have shown that a class of TGWs exists with wave fronts that have negative Gaussian curvature. We have investigated in some detail the properties of two such radiation fields. In particular, our study of the Jacobi equation in these spacetimes reveals a \emph{longitudinal} signature that is a distinct departure from transversality.  

In this first discussion of TGWs, many questions remain unanswered. For instance, how can TGWs be generated by realistic sources? Are there TGWs with wave fronts that have positive Gaussian curvature? Are there TGWs with a cosmological constant $\Lambda$?  Further investigation is necessary to tackle these problems.  

\appendix

\section{Field Equations with $\Lambda$}

The Einstein field equations in vacuum but with a cosmological constant $\Lambda$ can be expressed as
\beq \label{A1}
R_{\mu \nu} = \Lambda\, g_{\mu \nu}\,.
\eeq
We assume a metric of the form
\begin{equation}\label{A2}
 ds^2 = - e^{A(x, u)}(dt^2-dz^2) + e^{B(x, u)}\,dx^2 + e^{C(x, u)}\,dy^2\,,
\end{equation}
where $u=t-z$ is the retarded null coordinate. With metric~\eqref{A2}, field equations~\eqref{A1} can be reduced to the following five equations:
\beq \label{A3}
A_x^2+2\,A_x\,C_x +4\,\Lambda \,e^B=0\,,
\eeq
where $A_x:=\partial A/\partial x$, etc., 
\beq \label{A4}
2\,A_{xx} +A_x^2 -A_x(B_x+C_x)=0\,,
\eeq
\beq \label{A5}
2\,C_{xx} +C_x^2 -A_x^2 -B_x\,C_x=0\,,
\eeq
\beq \label{A6}
A_{xu} + C_{xu} = \frac{1}{2}\, A_x(B_u+C_u) + \frac{1}{2}\, C_x(B_u-C_u)\,,
\eeq
\beq \label{A7}
A_u\, (B_u+C_u)  = \frac{1}{2}\,(B_u^2+C_u^2) + B_{uu}+C_{uu}\,.
\eeq

It has not been possible to find an exact solution---that is, a solution that can be expressed using familiar functions---of these equations for $\Lambda \ne 0$. 

\section{Gaussian Curvature of the Wave Front}

Suppose that the metric of the wave front is given by
\begin{equation}\label{B1}
 d\sigma^2 = e^{P(x,y)} \,dx^2 + e^{Q(x,y)}\,dy^2\,.
\end{equation}
Then, the Gaussian curvature of this surface is given by
\begin{equation}\label{B2}
 K_G = -\frac{1}{4}\,e^{-P}\,[2\,Q_{xx} - (P_x- Q_x) Q_x] -\frac{1}{4}\,e^{-Q}\,[2\,P_{yy} +(P_y-Q_y) P_y]\,.
\end{equation}

As an application of this formula, consider the spacetime metric
\begin{equation}\label{B3}
 ds^2 = -dt^2 + dz^2 +u^2 \,(e^P\, dx^2 + e^Q\,dy^2)\,,
\end{equation}
which has the null Killing vector field $\partial_v = \partial_t + \partial_z$ and the homothetic vector field $t\,\partial_t + z\,\partial_z$. It represents \emph{flat} spacetime if the following field equation is satisfied
\begin{equation}\label{B4}
e^P\,[2\,P_{yy} +(P_y-Q_y) P_y] + e^Q\,[2\,Q_{xx} - (P_x- Q_x) Q_x] =0\,.
\end{equation}
That is, Eq.~\eqref{B4} implies that the Gaussian curvature of the wave front vanishes by Eq.~\eqref{B2}; moreover, the spacetime given by
\begin{equation}\label{B5}
 ds^2 = -dt^2 + dz^2 +u^2 \,(dx^2 + dy^2)\,
\end{equation}
is simply flat, cf.  the end of Appendix B of Ref.~\cite{Bini:2017qnd}.

\section{Scalar field on Harrison's TGW Background}

To distinguish TGWs from plane gravitational waves, one can explore the behavior of various perturbing fields on these backgrounds. As an example of this approach,  let us consider a massless scalar field on the  background of Harrison's solution~\eqref{G1}. We write the corresponding metric  in the form
\beq \label{C1}
ds^2 = -x^{4/3} du dv + \lambda^2 u^{6/5} dx^2 + x^{-2/3} u^{-2/5} dy^2,
\eeq
where we assume that $x \ge 0$ and $u \ne 0$. 
The massless scalar field equation is
\beq\label{C2}
\Box \Psi =g^{\mu\nu}\nabla_\mu (\partial_\nu \Psi)=0\,,
\eeq
which can be written as 
\beq\label{C3}
\frac{\partial}{\partial x^\mu}\left(\sqrt{-g}\,g^{\mu\nu}\,\frac{\partial \Psi}{\partial x^\nu}\right) = 0\,.
\eeq
With metric of the form~\eqref{C1}, we have $\sqrt{-g} = \lambda \,x\,u^{2/5} / 2$ and the nonzero components of $g^{\mu\nu}$ are given by $g^{uv} = g^{vu} = -2\,x^{-4/3}$, $g^{xx}= u^{-6/5}/\lambda^2$ and $g^{yy} = x^{2/3} \,u^{2/5}$. Writing
\beq\label{C4}
 \Psi = e^{i \,K_v\, v + i\,K_y\, y}\, \psi(x, u)\,,
\eeq
the massless scalar wave equation reduces to 
\beq\label{C5}
 (x\,\psi_x)_x - 2i \lambda^2 K_v \,x^{-1/3} u^{4/5} \,[u^{2/5}\,\psi_u + (u^{2/5}\psi)_u] - \lambda^2 K_y^2 \,x^{5/3} u^{8/5}\,\psi = 0\,.
\eeq
Here, we work with a complex amplitude $\Psi$. The wave equation is linear; therefore, the real and imaginary parts of $\Psi$ constitute real solutions of the massless scalar wave equation. 

It may not be possible to express the solution of this equation in terms of familiar functions. To illustrate this point, let us assume that $K_y = 0$ in Eq.~\eqref{C5}. We then look for a solution of the form 
\beq \label{C5a}
\psi(x, u) = \mathbb{X}(x)\,\mathbb{U}(u)\,. 
\eeq
It is then a simple matter to find $\mathbb{U}$,
\beq \label{C5b}
\mathbb{U}(u) = \mathbb{U}_0\,u^{-1/5} \,\exp{\left(-\frac{5i\mathbb{C}}{2 \lambda^2 K_v} \,u^{-1/5}\right)}\,, 
\eeq
where $K_v \ne 0$ and $\mathbb{U}_0$ as well as $\mathbb{C}$ are in general complex constants. However, the corresponding differential equation for $\mathbb{X}$ can be written as
\beq \label{C5c}
x^{1/3}\, \frac{d}{dx}\left(x\,\frac{d\mathbb{X}}{dx}\right) + 2\,\mathbb{C}\,\mathbb{X} = 0\,, 
\eeq
which does not appear to have a solution expressible in terms of familiar functions except
 for $\mathbb{C} = 0$.
 
It is a peculiar feature of Eq.~\eqref{C5} that solutions exist for $K_v = 0$ as well. In this case, Eq.~\eqref{C5} reduces to
\beq\label{C5d}
 (x\,\psi_x)_x  - \lambda^2 K_y^2 \,x^{5/3} u^{8/5}\,\psi = 0\,.
\eeq 
Let us now define $\xi$,
\beq \label{C5e}
 \xi := \frac{3}{4} \lambda \, K_y\, x^{4/3} u^{4/5}\,.
\eeq
Then, $\psi = \varphi(\xi)$ satisfies the modified Bessel equation of zero order, namely, 
\beq \label{C5f}
\varphi_{\xi\xi}+ \frac{1}{\xi} \varphi_\xi -\varphi = 0\,.
\eeq
 
It is interesting to compare this situation with the propagation of the massless scalar field on the background of a \emph{plane} gravitational wave with metric of the form
\beq \label{C6}
ds^2 = - du\, dv + \lambda^2 u^{6/5} dx^2 +  u^{-2/5} dy^2\,,
\eeq
where the metric coefficients can be obtained from those of  Eq.~\eqref{C1} with $x = 1$. This is a special case of the metrics discussed in Section V of Ref.~\cite{Bini:2017qnd} with $s_2 = 3/5$ and $s_3 = -1/5$.  In this case, the solution of the massless scalar field equation can be written as
\beq\label{C7}
 \Psi = e^{i \,K_v\, v + i\,K_x\,x + i\,K_y\, y}\, \phi(u)\,,
\eeq
where $\phi (u)$ is given by
\beq\label{C8}
 \phi (u) = \phi_0\,u^{-1/5} \,\exp{\left(-\frac{5 i K_x^2}{4 \lambda K_v} \,u^{-1/5} + \frac{5i K_y^2}{28 K_v}\,u^{7/5}\right)}\,.
\eeq
Here, $K_v \ne 0$ and $\phi_0$ is a constant of integration.

\section{Behavior of solutions of Eqs.~\eqref{J22}--\eqref{J23}}

The coefficients of the linear system~\eqref{J22}--\eqref{J23} are not all constants; therefore, the analysis of the system is expected to rely on qualitative or numerical methods. These are employed in this appendix. To this end, we transform Eqs.~\eqref{J22}--\eqref{J23} in the usual manner into four first-order linear differential equations involving $X^{\hat 1}$, $dX^{\hat 1}/dS$, $X^{\hat 3}$ and $dX^{\hat 3}/dS$.

Let us first discuss the long-time asymptotics,  that is the fate of solutions starting at $S=0$  as $S\to \infty$.  Based on numerical experiments, it appears that all solutions are bounded and have limits as $S\to \infty$. Moreover, the limiting behavior may be specified as an affine transformation from initial to final data such that
\begin{equation}\label{D1}
\lim_{S\to \infty}\left(X^{\hat 1}(S), \frac{dX^{\hat 1}}{dS}(S), \frac{dX^{\hat 3}}{dS}(S)\right) = (0,0,0)\,,
\end{equation}
while
\begin{equation}\label{D2}
\lim_{S\to \infty}X^{\hat 3}(S) = -k_1X^{\hat 1}(0)-k_2  \frac{dX^{\hat 1}}{dS}(0)+k_3 X^{\hat 3}(0)+k_4 \frac{dX^{\hat 3}}{dS}(0)+k_5\,,
\end{equation} 
where the $k_i$ are  positive constants with approximate values
\begin{equation}\label{D3}
k_1=0.572\,, \qquad k_2= 0.441\,, \qquad k_3=0.724\,,\qquad k_4=0.842\,, \qquad k_5=0.170\,.
\end{equation} 
One way to verify this result is to recall the basic fact from linear systems theory that the general solution is given by a particular solution plus a linear combination of fundamental solutions.  A convenient choice is the particular solution starting from zero initial data and the four fundamental solutions  corresponding to initial data given by the usual basis vectors of 4-dimensional Cartesian space. All that remains is numerical integration to obtain the limiting values of the particular and fundamental solutions. We find that  all solutions approach the $X^{\hat 3}$ direction with their final positions determined by the initial data. In contrast to the treatment of the Jacobi equation for the 
\emph{plane} GW~\eqref{H14a}, the initial $X^{\hat 3}$ component is never fixed during the evolution. However, its limiting value can be the same as its initial value by choosing the remaining components of the initial data so that the third component of the limit vector has the value $X^{\hat 3}(0)$. This happens on a hypersurface  in the space of initial data; thus, the probability of this event is zero.

The boundedness of solutions is of course implied by the boundedness of a particular solution and all the fundamental solutions. So, the numerical computations suggest that this is indeed the case.  For an analytic proof, one may use  the variation of parameters formula with respect to the constant part of the system matrix and then employ a Gronwall estimate~\cite{CH}.  More specifically and abstractly,  consider a vector ordinary differential equation,
\beq \label{D4} 
 \dot x = \tilde{A}\, x+\tilde{B}(t) \,x+\tilde{b}(t)\,,
 \eeq
where $\tilde{A}$  is a constant square matrix, $\tilde{B}$ is a continuous matrix function of time and $\tilde{b}(t)$ is a continuous vector function all defined  for $t\ge 0$. Using the usual Euclidean norm, suppose that there are positive constants $M$, $K$, $\tilde{\lambda}$, $\mu$ and $L$ such that for all $t\ge 0$ and all vectors $\tilde{v}$
\begin{equation}\label{D5}
 \abs{e^{t \tilde{A}}\tilde{v}}\le M \abs{\tilde{v}}, \qquad \abs{\tilde{B}(t) \tilde{v}}\le K e^{-\tilde{\lambda} t} \abs{\tilde{v}}, \qquad \abs{\tilde{b}(t)}\le Le^{-\mu t}\,.
\end{equation}
By variation of parameters, 
\beq \label{D6}
\abs{x(t)}\le \abs{e^{t \tilde{A}} x(0)}+\int_0^t \abs{e^{(t-s)\tilde{A}}\tilde{B}(s) x(s)}\,ds+\int_0^t \abs{e^{(t-s)\tilde{A}}\tilde{b}(s)}\,ds\,.
\eeq
Using the  inequalities~\eqref{D5}, 
\beq \label{D7}
\abs{x(t)}\le M\abs{ x(0)}+\int_0^t MKe^{-\tilde{\lambda} s} \abs{ x(s)}\,ds+ \frac{M L}{\mu} (1- e^{-\tilde{\lambda} t})\,.
\eeq
The last term is bounded by $ML/\mu$.  Gronwall's inequality implies
\beq \label{D8}
\abs{x(t)}\le (M+\frac{ML}{\mu}) e^{MK \int_0^t e^{-\tilde{\lambda} s} \,ds}\le  M\frac{\mu+ L}{\mu} e^{MK (1-e^{-\tilde{\lambda} t})/\tilde{\lambda}}
\le M\frac{\mu+ L}{\mu} e^{MK/\tilde{\lambda}}<\infty\,, 
\eeq
as required.

The hypotheses are true for system~\eqref{J22}--\eqref{J23} recast in the usual manner as a first-order linear system and  with the natural definitions of $\tilde{A}$, $\tilde{B}$ and $\tilde{b}$.  In particular, the four-dimensional constant-coefficient system matrix $\tilde{A}$ has four distinct eigenvalues: three negatives and one zero.  The fundamental matrix $\exp({t\tilde{A}})$ therefore propagates fundamental solutions starting in the corresponding eigenspaces of $\tilde{A}$, three of which decay in norm and one remains constant. Using linearity, the desired estimate $\abs{\exp({t \tilde{A}})\,\tilde{v}}\le M \abs{\tilde{v}}$ follows for an appropriate choice of constant $M$. Required estimates for $\tilde{B}$ and $\tilde{b}$ do not involve exponentiation and are straightforward.  

Similar arguments can be used to prove the conjecture---based on numerical experiments---that $X^{\hat 1}(T)$ is not only  bounded on the interval $0< T\le 1$, but in fact $X^{\hat 1}(T)\to 0$ as $T\to 0^+$; or equivalently, $X^{\hat 1}(S)\to 0$ as $S\to \infty$.  Our proof to follow also shows that the derivatives  $dX^{\hat 1}/dS$ and $dX^{\hat 3}/dS$ converge to zero exponentially fast as $S\to \infty$.  Unfortunately, our argument proves only an upper bound for their rate of convergence to zero. A lower bound would be needed to conclude  that both $dX^{\hat 1}/dT$ and $dX^{\hat 3}/dT$  blow up  in absolute value as $T\to 0^+$, a conjecture that is also supported by numerical experiments. To see the problem, recall for example that by the change of variables 
$dX^{\hat 1}/dT =-\exp(S)\, dX^{\hat 1}/dS$. This indeterminate form, which is expected to blow up, would  do so  if $\abs{dX^{\hat 1}/dS}$ does not approach zero too rapidly. 

Our proof requires several steps that are merely outlined here for the sake of brevity.  The underlying idea is to split off the part of the constant-coefficient system matrix $\tilde{A}$ corresponding to its zero eigenvalue and thus to take advantage of the exponential decay afforded by the remaining three eigenvalues: $-1$, $-3/5$ and $-2/5$.  The boundedness of solutions already proved is also a key ingredient. 

To accomplish the desired splitting, note that $\tilde{A}$ is diagonalizable.  In fact, taking $\tilde{Q}$ to be the matrix whose columns are eigenvectors corresponding to the listed eigenvalues and the remaining zero eigenvalue in that order, $D:=\tilde{Q}^{-1}\tilde{A} \tilde{Q}$ is diagonal with the eigenvalues in the specified order along the main diagonal. With the change of variables $x=\tilde{Q}\, y$ and using $t$ instead of $S$, the first-order linear system is recast in the form
\beq \label{D9}
\dot y=D y+\tilde{Q}^{-1}\tilde{B}(t)\tilde{Q} y+\tilde{Q}^{-1} \tilde{b}(t)\,.
\eeq
The first three differential equations in this vector system may be recast in the vector form
\beq \label{D10} 
\dot z=D_1z+ \tilde{B}_1(t) z+ y_4 \tilde{B}_2(t) + \tilde{b}_1(t)\,,
\eeq
where the coupling to the fourth differential equation is through the term $y_4 \tilde{B}_2(t)$ for the $3\times 1$ matrix $\tilde{B}_2(t)$. This latter matrix is simply the first three components of the last column of $\tilde{Q}^{-1}\tilde{B}(t)\tilde{Q}$ and of course $\tilde{B}_1(t)$ is its upper $3\times 3$ diagonal block.  As in the proof of boundedness, there are positive constants $L$ and $K$ and exponential estimates given by 
\beq \label{D11}
\abs{e^{t D_1} \tilde{v}}\le e^{-2t /5}\abs{\tilde{v}}\, 
\eeq
and
\beq \label{D12}
\abs{\tilde{B}_1(t) \tilde{v}}\le L e^{-2 t/5}\abs{\tilde{v}}\,, \qquad \abs{\tilde{B}_2(t) \tilde{v}}\le L e^{-2 t/5}\abs{\tilde{v}}\,, \qquad
\abs{\tilde{b}_1(t)} \le K  e^{-7 t/5}\,.
\eeq

Again the variation of parameters formula is employed, the triangle law estimate is made,  the boundedness of $y_4$ is noted as a corollary of our previous result, and all  exponential estimates are used.  After some manipulation and obvious estimates,  we find that
\beq \label{D13}
e^{2 t/5}\abs{z(t)}\le \abs{z(0)}+L t+1+\int_0^t e^{-2 s/5} (e^{2s/5} \abs{z(s)}) \,ds\,.
\eeq
Gronwall's inequality followed by a simple integral estimate implies
\beq \label{D14}
\abs{z(t)} \le (\abs{z(0)}+L t+1) e^{5/2} e^{-2 t/5}\,.
\eeq
In particular, $z(t)$ converges to zero as $t\to \infty$. This is enough, after some interpretation,  to obtain the claimed limit 
$X^{\hat 1}(S)\to 0$ as $S\to \infty$, or, equivalently, $X^{\hat 1}(T)\to 0$ as $T\to 0^+$.

Further discussion of system~\eqref{J16}--\eqref{J18} is contained in Ref. \cite{ref25}.

\section*{Acknowledgments}

We are grateful to Marcello Ortaggio and Kjell Rosquist for their insightful comments on our manuscript. D.B. thanks the Italian INFN (Naples) for partial support.

\end{document}